\documentclass[aps,twocolumn,showpacs,preprintnumbers,nofootinbib,prd,superscriptaddress,groupedaddress,10pt]{revtex4-2}
\pdfoutput=1

\makeatletter
\def\l@subsubsection#1#2{}
\def\l@subsubsubsection#1#2{}
\makeatother

\usepackage{graphicx,amssymb,amsmath,amsthm,amsfonts,epsfig,epsf,fixmath}
\usepackage[usenames]{color}
\usepackage{epstopdf}
\usepackage{mathtools}
\usepackage{aas_macros}
\usepackage{bm}
\usepackage{dcolumn}
\usepackage{latexsym}
\usepackage{rotating}
\usepackage{longtable}

\usepackage{enumerate}
\usepackage{tensor,multirow}
\usepackage{url}
\usepackage[linktocpage]{hyperref}

\def\nn{\nonumber}

\newcommand{\pmat}{\begin{pmatrix}}
\newcommand{\fpmat}{\end{pmatrix}}
\newcommand{\eq}{\begin{equation}}
\newcommand{\feq}{\end{equation}}
\newcommand{\cas}{\begin{cases}}
\newcommand{\fcas}{\end{cases}}

\newcommand{\eqarray}{\begin{eqnarray}}
\newcommand{\feqarray}{\end{eqnarray}}

\newcommand{\be}{\beta}

\def\be{\begin{equation}}
\def\ee{\end{equation}}
\def\bea{\begin{eqnarray}}
\def\eea{\end{eqnarray}}

\newcommand{\ft}[2]{{\textstyle\frac{#1}{#2}}}

\begin{document}
\title{QNMs of branes, BHs and fuzzballs from Quantum SW geometries}
\author{
Massimo Bianchi,
Dario Consoli,
Alfredo Grillo,
Jos\`e Francisco Morales,
}

\affiliation{Dipartimento di Fisica,  Università di Roma ``Tor Vergata"  \& Sezione INFN Roma2, Via della ricerca scientifica 1, 
00133, Roma, Italy}

\begin{abstract}
 QNMs govern the  linear response  to perturbations of BHs, D-branes and fuzzballs  and  the gravitational wave signals in the ring-down phase of binary mergers. A remarkable connection between QNMs of neutral BHs in 4d and quantum SW geometries describing the dynamics of ${\cal N}=2$ SYM theories has been recently put forward. We extend the gauge/gravity dictionary to a large class of gravity backgrounds including charged and rotating BHs of Einstein-Maxwell theory in $d=4,5$ dimensions, D3-branes, D1D5 `circular' fuzzballs and  smooth horizonless geometries; all related to ${\cal N}=2$ SYM with a single $SU(2)$ gauge group and fundamental matter. We find that photon-spheres, a common feature of all examples, are associated to degenerations of the classical elliptic SW geometry whereby a cycle pinches to zero size. Quantum effects resolve the singular geometry  and lead to a spectrum of quantized energies, labeled by the overtone number $n$.  We compute the spectrum of QNMs using exact WKB quantization, geodetic motion and numerical simulations and show excellent agreement between the three methods.   We explicitly illustrate our findings for the case D3-brane QNMs.  

\end{abstract}

\maketitle 


\section{Introduction} 
 
Direct detection of gravitational waves (GWs) produced in binary mergers allows to test General Relativity (GR) in extreme (e.g. strong-field) regimes and to discriminate Black Holes (BHs) from fuzzballs or other exotic compact objects (ECOs) in terms of their multipoles \cite{Bianchi:2020bxa,Bianchi:2020miz,Bena:2020see,Bena:2020uup,Bah:2021jno}, shadows \cite{Bianchi:2020des,Bacchini:2021fig}, or tidal effects \cite{Martinec:2020cml,Bena:2020iyw}.

The GW signal can be decomposed into three main phases: inspiral, merger and  
ring-down. The latter is dominated by the quasi-normal modes (QNMs) present in the linear response to perturbations. In the case of ECOs in alternative theories of gravity and fuzzballs in string theory \cite{Mathur:2009hf}, the `prompt ring-down signal' decomposes in the early-stage resonant modes,  produced around their `photon-spheres' that may differ from the BH ones. At later stages both ECOs and fuzzballs produce a peculiar train of echoes, probing their internal `cavity'  and not only their external `walls', with significant deviations from GR \cite{Cardoso:2017cqb,Correia:2018apm,Cardoso:2016rao,Mayerson:2020tpn}. 

In the eikonal approximation, the real and imaginary parts of the QNM frequencies or the `prompt ring-down modes' can be expressed as \cite{Cardoso:2008bp,Mashhoon:1985cya,Schutz:1985km,Iyer:1986np,Bianchi:2020des,Bianchi:2020yzr,ToVSapQNM}
 \be
\omega_{QNM} \approx {{\omega}}_c(\ell)  - i (2n+1)\lambda_c. \label{wgeo}
 \ee
where $\omega_c(\ell)$ is the orbital frequency of unstable `circular' orbits forming the light-ring, while $\lambda_c$ is the Lyapunov exponent governing the chaotic behaviour of nearly critical geodesics around it.

The crucial role played by QNMs in discriminating BHs from fuzzballs or other ECOs motivated renewed effort in their determination with higher and higher accuracy \cite{Bena:2020yii,ToVSapQNM}. Going beyond the WKB approximation proved to be a hard task even in the simpler GR context. QNMs of Schwarzschild BHs are governed by the celebrated Regge-Wheeler-Zerilli equation \cite{RegWheel,Zerilli:1970se} that can be put in Schr\"odinger-like (canonical) form
\be
\Psi''(z) +Q(z) \Psi(z)=0 \label{waveeq}
\ee
where $Q(z)$ is a rational function with poles associate to horizons or singularities and zeros to turning points in the $\hbar {\,\to\,} 0$ limit. The problem is classically integrable but the wave equation is not exactly solvable and the QNMs are only known numerically, with high precision, though.

The Kerr BH case is even more involved in that the two equations for radial and angular motion, known as Teukolsky equations \cite{Teukolsky:1972my,Dias:2015wqa,Pani:2013ija,Pani:2013wsa,Mark:2014aja}, talk to each other via the `separation constant', the `wavy' analogue of Carter constant.

A very efficient approach was later developed by Leaver \cite{Leaver:1985ax,Leaver:1990zz} that allowed to determine numerically the QNMs of Kerr or Reissner-Nordst\"om BHs by making use of continuous fractions.

More recently a remarkable connection with quantum Seiberg-Witten (SW) curves of ${\cal N}{\,=\,}2$ super-symmetric Yang-Mills (SYM) theory on a Nekrasov-Shatashvili (NS) $\Omega$-background ($\epsilon_1{\,=\,}\hbar$, $\epsilon_2{\,=\,}0$) \cite{Seiberg:1994rs,Seiberg:1994aj,Matone:1995rx,Nekrasov:2009rc,Alday:2009aq} was revealed in \cite{Grassi}. In the NS background, the gauge theory is described by a differential equation of type (\ref{waveeq}), that  is solved (exactly in $\hbar$) by the NS prepotential \cite{Mironov:2009uv,Zenkevich:2011zx,Bourgine:2017aqk,Fioravanti:2019vxi,Grassi:2018bci,Grassi:2019coc}. QNMs of neutral rotating BHs in four dimensions were computed  by imposing exact WKB quantization conditions on the quantum periods of the SW curve. Possible extensions to BHs in AdS or in higher dimension were sketched in \cite{Grassi}, while finite frequency greybody factor, QNMs and Love
numbers of Kerr BHs were determined using irregular 2-d conformal blocks \cite{BonTanzetc}.

Aim of the present investigation is to extend the gauge/gravity dictionary to a large class of gravity backgrounds. The common feature of the examples we consider is the existence of a photon-sphere, that traps light on unstable null orbits eventually leaking out in ring-down modes. We find that QNMs are encoded in differential equations of type (\ref{waveeq}) with $Q_{SW}(y)=P_4(y)/P_2(y)^2$ given as a ratio of polynomials of degree 4 in the SW variable $y$.

We find that photon-spheres are associated to degeneration of the quantum elliptic SW geometry where a cycle pinches to zero size in the semiclassical eikonal limit $\hbar \to 0$. QNMs can be obtained by solving the exact WKB quantization conditions
\be
\label{aadquantum}
 a_\gamma=  \oint_{\gamma}  \lambda_{SW} =  \hbar  (n+\ft12)
\ee
with $\gamma$ the cycle vanishing at the photon-sphere and $\lambda_{SW} $ the quantum  SW differential
\be
\lambda_{SW}=
{\hbar \over 2\pi} \sqrt{Q_{SW}(y)} dy +\ldots \label{lambdaclas}
\ee
with dots denoting higher $\hbar$-corrections.
The singular geometry is resolved by quantum effects leading to a spectrum of quantized energies labeled by the overtone number $n$.

We restrict ourselves to classical integrable geometries that allow to write down separate, but some times intertwined, ordinary differential equations, generalising the famous Teukolsky equations. 
We find that radial and angular equations can be mapped to ${\cal N}=2$ theories with a single $SU(2)$ gauge group factor, possibly with hypermultiplets in the fundamental.
The dictionary is not one-to one: a single gravity background can be described in terms of apparently unrelated gauge theories with different number of flavours related by modular transformations of the underlying elliptic geometry.

The systems we will consider include Kerr-Newman BHs in $D=4$,
CCLP geometries describing charged and rotating solutions of Einstein-Maxwell theory in $D=5$ \cite{CCLP1,CCLP2}, including Myers-Perry \cite{MyerPer} and BMPV BHs \cite{BMPV},  D3-branes and D1D5 circular fuzzballs \cite{Lunin:2001fv}.


For simplicity, we will only consider neutral scalar perturbations. Metric and vector perturbations lead to similar equations but the derivation is more laborious in general, and may spoil the elegance of the approach \cite{Dias:2015wqa,Pani:2013ija,Pani:2013wsa,Mark:2014aja}.

One immediate outcome of our analysis is a better knowledge of their QNMs that play a crucial role in (in)stability analyses in these contexts \cite{Cardoso:2005gj,Eperon:2016cdd,Chakrabarty:2019ujg,Bianchi:2019lmi,Maggio:2018ivz}.


The presentation is organised as follows. We describe the three different approaches to study QNMs: geodetic motion, SW exact quantization, numerical methods based on continuous fractions.   
We illustrate the procedure for D3-branes as working example and present the gauge/gravity dictionary for a large number of BHs and D-brane gravity solutions. Detailed solutions of the various problems will be described in a forthcoming paper.

 \section{Geodetic Motion}
 
 Geodetic motion of massless neutral probes is governed by the null Hamiltonian ${\cal H}=\ft12 g^{MN} P_M P_N=0$. If the dynamics is separable, radial and angular motion can be disentangled and effectively described by one dimensional Hamiltonians. 

For example, for a D3-brane the metric reads
\be
\label{sphermetric}
ds^2 = H(r)^{-{1\over 2}}  (-dt^2+d{\bf x}^2)  + H(r)^{{1\over 2}}   (dr^2 + r^2 d\Omega_5^2) 
\ee   
where ${\bf x}$ are the longitudinal coordinates, $H(r)=(1+\ft{L^4}{r^4} )$  and $d\Omega_5^2$ denotes the metric of the transverse round $S^5$-sphere. Radial motion is governed by the null Hamiltonian
${\cal H} \sim P_r^2 - Q_{\rm geo} (r)$ with $P_r$ the radial momentum and
\be 
Q_{\rm geo}(r) ={  \omega^2\,  (r^4+L^4)- J^2 r^2  \over  r^4 }  
\ee
where $\omega^2=E^2-{\bf k}^2$ and $J$ is the transverse angular momentum. 
  Simple zeros $r_*$ of  $Q_{\rm geo}(r)$ are associated to turning points, while double zeros $r_c$ define the photon-sphere, where light gets trapped orbiting around forever for specific choice of the frequency ${{\omega}}_c$. The critical equations
 \be
 Q_{\rm geo}({{\omega}}_c,r_c)=\partial_r Q_{\rm geo}({{\omega}}_c, r_c)=0 \label{qqp}
 \ee 
can be solved for $r_c$ and ${{\omega}}_c$. Nearly critical geodesics fall with radial velocity $\dot{r} \approx   - 2 \lambda_c (r-r_c)$, where 
\be
 \lambda_c =\left(  \sqrt{2}\, \partial_\omega Q_{\rm geo} (r_c,{{\omega}}_c)\right)^{-1} \sqrt{ \partial_r^2 Q_{\rm geo}(r_c,{{\omega}}_c)} \label{lyapunov}
\ee 
is the Lyapunov exponent that characterises the chaotic behaviour of geodesics around the photon-sphere. 
For D3-branes one finds
\be\label{photonsp}
   {\rm D3:} \qquad r_c=L \quad, \quad {{\omega}}_c \approx {{\ell + 2} \over \sqrt{2} L} \quad, \quad \lambda_c={1\over 2L}   
\ee
where we used $J\approx \sqrt{\ell(\ell+4)} \approx \ell+2$ in the large $\ell$ limit.

\section{Wave equation}

 The wave equation for a massless scalar field (or a scalar fluctuation of the metric) in the metric $g_{MN}$ reads 
\be
\square \, \Phi = 
 g^{-\frac{1}{2}} \partial_M ( g^{\frac{1}{2}} g^{MN} \partial_N ) \,\Phi=0
\ee
For the metric (\ref{sphermetric}), using the ansatz
 \be
\Phi  = e^{-i E t + i {\bf k}{\cdot}{\bf x}}  r^{5/2} \Psi(r){\cal Y}^\ell_{\vec m}(\vec\theta)
\ee
with $\nabla_{S^5} {\cal Y}^\ell_{\vec m} =-\ell(\ell{+}4) {\cal Y}^\ell_{\vec m}$, one finds a radial equation in the canonical form (\ref{waveeq}) 
with
 \be
 Q(r)= {  4  \omega^2( r^4+L^4) -r^2 (4 \ell(\ell+4) +15) \over 4 r^4}\
 \label{qhbar}
 \ee
 QNMs correspond to solutions describing outgoing waves at infinity
 \be
\Psi(r) \sim e^{i\omega r} \quad {\rm as} \quad r\rightarrow +\infty \label{bc_inf}
\ee
and in-going waves in the deep interior of the photon-sphere (e.g. at the horizon) or vanishing at the centrifugal barrier of a smooth horizonless compact object).  For D3-branes, one requires
 \be
  {\rm D3:} \qquad \Psi(r) \sim e^{i {\omega L^2 \over  r} } \quad {\rm as} \quad r\rightarrow 0 \label{bc}
\ee
   In the semiclassical limit, where $\omega $, $\ell$ are large, the equation can be integrated and QNMs  follow from the Bohr-Sommerfeld quantization condition
 \be
\int_{r_-}^{r_+}   \sqrt{Q(r)}  \, dr={\rm } \pi \left( n+ \ft12 \right)  \label{bohr}
\ee
 with   $r_\pm$ the inversion points, where $Q(r_\pm)=0$.  The integral (\ref{bohr}) can be approximated by expanding $Q(r)$ around
   its minimum at $r_c$ to quadratic order leading to 
  \be
  {Q(r_c) \over \sqrt{2 \,\partial_r^2 Q(r_c) }}=- {\rm i}  \left( n+ \ft12 \right)  \label{wkb-radial}
\ee
 This equation can be solved by giving a small imaginary part to $\omega$, i.e. writing $\omega=\omega_R+{\rm i}\, \omega_I$ and solving perturbatively
 in $\omega_I$. To linear order in $\omega_I$ one finds
 \be
 {\rm D3:}  \quad \omega  \approx {1 \over \sqrt{2} L}\sqrt{ (\ell+2)^2-\ft14 } -{{\rm i} \over 2L} (2n+1)  
 \label{WKBapprox}
 \ee
  in agreement with the geodetic results \eqref{wgeo}, \eqref{photonsp} at large $\ell$.  
  
%
 
\section{QNMs from quantum SW curves }
 
%
 The dynamics of ${\cal N}{\,=\,}2$ gauge theory with fundamental matter in a non-trivial NS-background can be described by the differential equation  
  \be
\left[ q\,  y^{1\over 2} \, P_+(x) \,  y^{1\over 2} +P_0(x)+\, y^{-{1\over 2}} \, P_-(x) \, y^{-{1\over 2} }     \right]\Psi=0 \label{diffsw}
 \ee
 where 
\bea
&P_+(x) = \prod_{i=1}^{N_+} (x{-}m_{ i})~,~ P_-(x)=\prod_{i=N_++1}^{N_f} (x{-}m_{ i}) \nn 
\\
&P_0(x) = x^2 {+}  q \delta_{N_f,3} \, x - \hat{u}  \label{ppms}
\eea
with $\hat{u}= u{+} q(m_1{+}m_2{+}m_3-\tfrac{\hbar}{2})\delta_{N_f,3}{-}q \delta_{N_f,2}$.
   Here $u=\ft12\langle {\rm tr} \varphi^2 \rangle=a^2+\ldots$ parametrizes the Coulomb branch, $m_i$ the masses, $q=\Lambda^{4-N_f}$ the gauge coupling. Finally  $x$, $y$ are operators satisfying the commutation relation $[x,\ln y]=\hbar$ and $\Psi$ an energy eigenstate. 
  
  One can view (\ref{diffsw}) as an ordinary differential equation  of second order  in $y$ by setting $\hat x=\hbar y \partial_y$  or as a difference
  equation in $x$ by setting  $y=e^{-\hbar \partial_x}$. 
    For instance, using $ P(x)y =y P(x+\hbar) $ to bring all the dependence on $y$ to the left, one can write (\ref{diffsw})  as
 \bea
0&=& \left[   A(y) \hat{x}^2 +  B (y)\,  \hat{x} +C (y) \right] \Psi(y)  \label{swcurve}\\
&=& \left[ q\,  y^2 \, P_+( \hat{x}+\ft{\hbar}{2})   + \, y P_0(\hat{x} )+\,  P_-(\hat x-\ft{\hbar}{2} ) \right]\Psi(y)\nn
\eea
 with $A(y)$, $B(y)$, $C(y)$ some polynomials of order at most two. For example, for $(N_+,N_-)=(1,2)$, one finds
 \bea
 A(y) &=& 1{+} y \quad, \quad B=q y(y{+}1){-}m_2{-}m_3{-}\hbar \nn\\
 C(y) &=& q y^2(\ft{\hbar}{2} {-}m_1) {-}y( \hat{u} +\hbar q ){+}(m_2{+} \ft{\hbar}{2})(m_3{+} \ft{\hbar}{2})~~~~
 \eea
  Massive fundamentals can be decoupled by sending $m\to \infty$ and $q\to 0$ keeping $\hat{q}=-q m$ fixed.
  Bringing equation (\ref{swcurve}) to canonical form one finds
  \be
Q_{SW}(y)  = {4 \,C \,A {-} B^2{+}2\, \hbar\, y(B\, A'{-}A\, B'){+}\hbar^2\, A^2  \over  4\, \hbar^2\, y^2\, A^2 } \label{qsw}
\ee
Alternatively, viewing (\ref{diffsw}) as a difference equation one can write the $\hbar$-deformed SW equation \cite{Poghossian:2010pn,Fucito:2011pn}
\be
     q\, M(x) \, W(x) W(x-\hbar)  + P_0(x) W(x)  +1  =0 \label{dswcurve0}
\ee
with $M(x)=P_+(x-\ft{\hbar}{2})P_-(x-\ft{\hbar}{2})$
and
\be
 W(x) ={1\over P_-(x+\ft{\hbar}{2}) }{ \hat\Psi(x)  \over \hat\Psi(x+\hbar) }  
\ee
  Eq. \eqref{dswcurve0} can be recursively solved order by order in $q$, $W(x)=-{1\over P_0(x)}+\ldots$. 
  The quantum period $a(u,q,\hbar)$ can therefore be written as a sum over residues
\be
\label{aperiod}
  a(u,q,\hbar) = \oint_{\alpha}  \lambda  = 2\pi {\rm i} \sum_{s=0}^\infty {\rm Res}_{\sqrt{u}+s \hbar} \lambda_x(x)    
\ee
of the $\hbar$-deformed Seiberg-Witten differential $\lambda_{SW}$
\be
 2\pi {\rm i}\, \lambda_{SW}(x) =-\,x\, d \ln  W(x)   
\ee
The $a_D$ period is computed in terms of  NS prepotential ${\cal F}(a,q,\hbar)$ via the identification
   \be
  2\pi {\rm i}\,a_D (a,q,\hbar) =  \partial_a {\cal F}(a,q,\hbar) \label{adperiod}
  \ee
  The prepotential can be determined by inverting $a(u)$ given in (\ref{aperiod}) for $u(a)=a^2+\ldots $ order by order in $q$, using the quantum version of the Matone relation $u(a,q,\hbar)=  q\, \partial_q {\cal F}(a,q,\hbar)$ \cite{Flume:2004rp} and adding the $q$-independent one-loop term.   

 The QNM frequencies are obtained by imposing the exact WKB conditions on the vanishing cycle and using  the gauge/gravity dictionary following from
 \be
  Q(y)=Q(z) \, z'(y)^2 +{z'''(y) \over 2 z'(y) } -\frac{3}{4} \left[ { z''(y) \over  z'(y) } \label{qqschwarzian}
   \right]^2
  \ee
 Using this dictionary, the WKB conditions  translate into an equation for the frequencies $\omega$ that can be solved numerically.

 \section{The D3-brane quasi-normal modes}
 
The radial wave equation for a scalar field (such as the dilaton) on a D3-brane background  can be mapped to the quantum SW curve of pure $SU(2)$ gauge theory or equivalently to  the Mathieu equation. The characteristic functions are
   \bea
Q_{SW}(y) &=& {4\, y^2  \, q+y(\hbar^2-4u)+4 \over 4\, \hbar^2 \,y^3} \nn\\
Q_{\rm Mathieu} (z) &=& \alpha-2 \beta\, \cos(2z) 
   \eea
and gauge/gravity/Mathieu parameters are related by
  \bea
 \alpha &=& {4 u \over \hbar^2}  = (\ell+2)^2 ~ ,~ \beta= { 4\sqrt{q} \over \hbar^2}  =\omega^2\, L^2   \nn\\
 r &=& {  \hbar\omega  L^2  \over 2} \sqrt{y}=L\, e^{{\rm i} z} \label{abdict}
\eea
The general solution to Mathieu equation reads
\be
\Psi(z)=c_1 \, {\rm me}(\alpha,\beta,z)+ c_2 \, {\rm me}(\alpha,\beta,-z)\label{psimat}
\ee
with ${\rm me}(\alpha,\beta,z)$ the exponential Mathieu function which is quasi-periodic 
${\rm me}(\alpha,\beta,z{+}\pi)=e^{\pi {\rm i} \nu} {\rm me}(\alpha,\beta,z)$, with $\nu(\alpha,\beta)$ the Floquet exponent.
The latter can be related to the quantum $a$-period at weak coupling and to the quantum $a_D$-period at strong coupling
\cite{He:2010if}. In the weak coupling limit, one finds \cite{Grassi:2019coc}
 \be
 a (u,q) = {\hbar   \over 2} \nu(\alpha,\beta)= {\hbar \over 2\pi {\rm i}} \log\left[ { {\rm me}(\alpha, \beta,\ft{\pi}{2})\over {\rm me}(\alpha, \beta,-\ft{\pi}{2}) }\right]
    \ee
 that matches (\ref{aperiod}) after using the  
 the weak coupling  ($\beta\ll \alpha$) expansion of the Mathieu function 
\bea
 {\rm me}(\alpha, \beta,z)&=& e^{i z \nu } \left[ 1{-}{\beta \over 4}  \left( {e^{ 2 i z} \over \nu{+}1} {-}{e^{-2 i z} \over \nu{-}1} \right) 
 +\ldots \right] \label{me}
 \eea
 and of its energy eigenvalue 
   \be
   \alpha =\nu^2{+}{\beta^2\over 2 (\nu^2{-}1) } {+}{(5\nu^2{+}7)\beta^4\over 32(\nu^2{-}1)^3(\nu^2{-}4)}  +\ldots 
   \label{ruleua}
   \ee
 The photon-sphere (\ref{photonsp}) corresponds to the opposite limit $\alpha \approx \pm 2\beta$ ($u \approx \pm 2 \sqrt{q}$) where
 the gauge theory is strongly coupled and the $a_D$-cycle degenerates.  In this limit, the energy eigenvalue 
 is given by \cite{NIST}
 {\small
 \be
\alpha = {-} 2\beta {+} 2s\, \sqrt{\beta} {-} {1\over 8} (1+s^2) {-} {s (s^2+3) \over 2^7 \sqrt{\beta}} {-} {5 s^4 +34 s^2 + 9\over 2^{12} \beta} {+} \ldots
\label{alphastrong}
\ee
 }
 with $s=2\nu$ and  the Floquet exponent $\nu$ is now related to the $a_D$-quantum period \cite{He:2010if}. The WKB quantization conditions translate then to $s=n$ with $n$ the overtone.  QNM frequencies are obtained by plugging (\ref{abdict}) into (\ref{alphastrong}) and solving numerically for $\omega$ as a function of $\ell$ and $n$. 

 The QNM wave function is obtained by imposing the boundary conditions (\ref{bc_inf}) and (\ref{bc}) at $z=\pm {\rm i } \infty$. For $n$ odd, the result can be written in terms of the Mathieu function $M^{(3)}_\nu( {\rm i} z,\beta)$ \cite{NIST}
\be
\psi(z) \sim {e^{2{\rm i} \sqrt{\beta}\cos z } \over (\cos z + 1)^{1/2}} 
\sum_{m=0}^\infty  {(-)^m a_m(\alpha, \beta) \over [4 i \sqrt{\beta} (\cos z + 1)]^{m}} 
\ee
with $z$ purely imaginary and the coefficients $D_m$ determined by the recursion relation with $D_{-1}=0$, $D_{0}=1$
\bea
&& (m{+}1)D_{m+1} {+} [(m{+}\ft12)^2 {+} 8{\rm i}\sqrt{\beta} (m{+}\ft14) {+} 2\beta - \alpha] D_m \nn\\
&&\qquad~~~~~~~ + 8{\rm i}\sqrt{\beta} m(m{-}\ft12 )D_{m-1}=0
\eea

\begin{figure*}[t]
\includegraphics[width=0.35\textwidth]{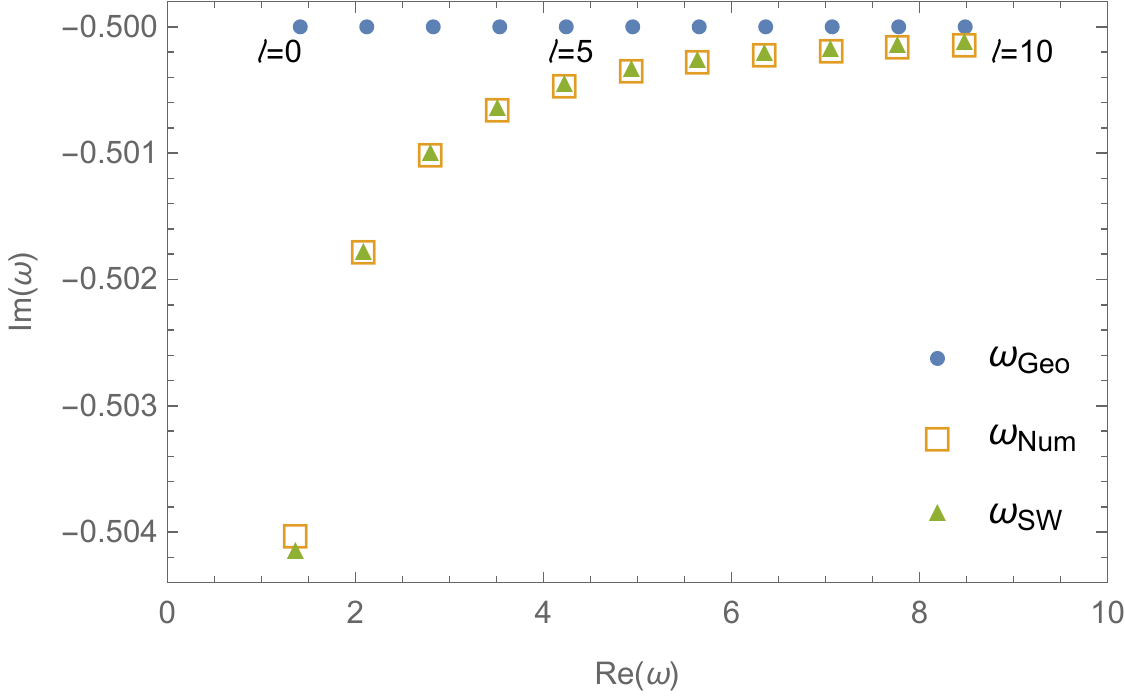}
\qquad\qquad\qquad\qquad\qquad
\includegraphics[width=0.35\textwidth]{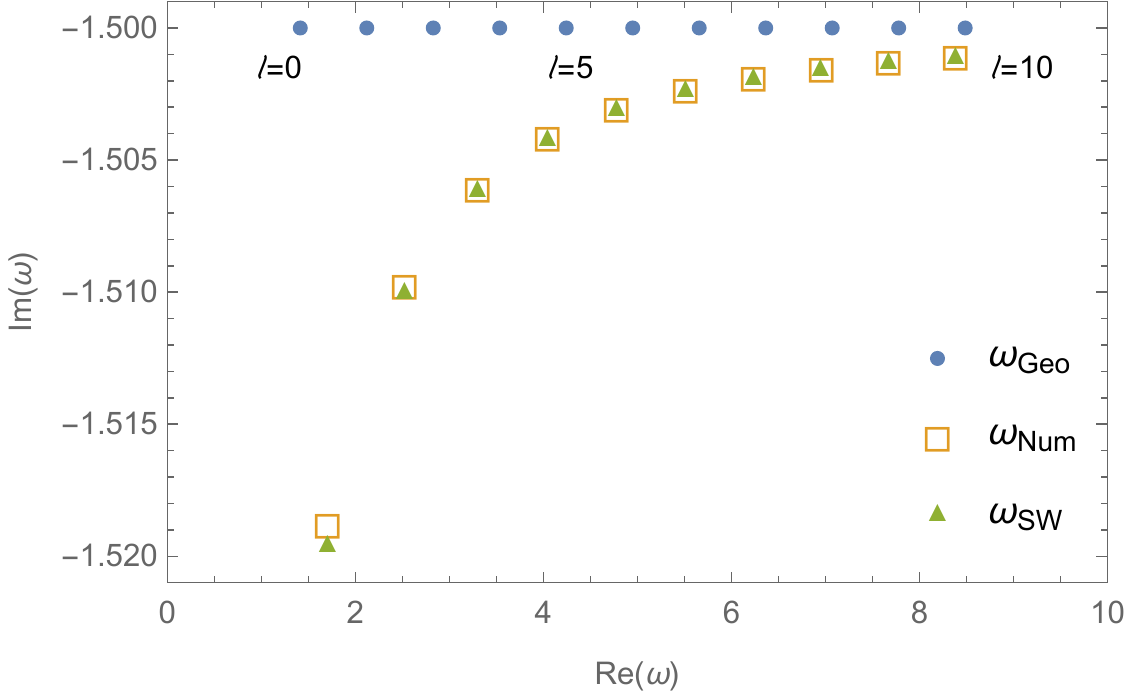}
\caption{QNMs of D3 branes for $n=0$ (left) and $n=1$ (right) for $\ell$ varying from 0 to 10.}
\label{fig-QNM-D3}
\end{figure*}

\section{Numerical computations} 

%
 To test the results obtained from geodetic motion and SW quantization one can solve the differential equation numerically using the method of continuous fractions introduced by Leaver \cite{Leaver:1985ax}.
 We  start from the ansatz
\begin{equation}
\Phi(z) = e^{ {\rm i}\omega z}(z{-}z_-)^{\sigma_-}(z{-}z_+)^{\sigma_+}\sum_{n=0}^\infty c_n \left(\frac{z{-}z_+}{z{-}z_-}\right)^n \label{ansatz}
\end{equation}
where $z_+ > z_-$. The constants $\sigma_+$, $\omega$ are determined by requiring that the ansatz solves the differential equation near $z_+$ and infinity, and imposing that only outgoing waves are present at infinity. 
On the other hand $\sigma_-$  is fixed by requiring that the recursion involve only three terms. One finds
\bea
\omega^2 &=& {P^{{}_{(IV)}}_4(z_+)\over 4!} \quad,\quad  \sigma_+(\sigma_+ {-}1) {+} \frac{P_4(z_+)}{\Delta^2} =0 \nn\\
\sigma_- &=& -\sigma_+ - {\rm i} \, \omega\, \Delta+  { {\rm i} \, P^{'''}_4(z_+)\over 12 \, \omega} 
\eea
with $\Delta=z_+ - z_-$ and prime denoting derivatives wrt $z$. There are two solutions for $\sigma_+$, generating two infinite towers of QNMs. Plugging the ansatz into the wave equation one finds the
recurrence 
\begin{equation}
\begin{aligned} \label{abeq}
& \alpha_n\, c_{n+1} + \beta_n\,  c_n + \gamma_n\, c_{n-1} = 0 
\end{aligned}
\end{equation}
with $c_{-1}=0$ and 
\bea
\alpha _n &=& (n{+}\sigma_+) (n{+}1{+}\sigma_+)+\frac{P_4(z_+) }{\Delta^2}\\
\beta _n &=&2 \left(n{+}\sigma _+\right) \left(\Delta  \nu {-}n{+}\sigma _-\right)+\frac{P_4'\left(z_+\right)}{\Delta}{-}\frac{2 P_4\left(z_+\right)}{\Delta ^2}\nn\\
\gamma _n&=&2 \Delta\, \nu  \left(\sigma _-+\sigma _+\right)+\left(n-\sigma _--1\right) \left(n-\sigma _-\right)\nn\\
&&{+}
\Delta ^2 \nu ^2+\frac{P_4\left(z_+\right)}{\Delta ^2}-\frac{P_4'\left(z_+\right)}{\Delta }+\frac{1}{2} P_4''\left(z_+\right)\nn
\eea
The QNM frequencies $\omega_n$ associated to the overtone $n$ can be obtained by truncating the recursion to level $N$ and solving 
numerically the
equation
\begin{equation}
\label{continuous_fraction}
\beta_n = \frac{\alpha_{n-1} \, \gamma_n}{\beta_{n-1} - \frac{\alpha_{n-2}\, \gamma_{n-1}}{\beta_{n-2}-\ldots}} +\frac{\alpha_n\gamma_{n+1}}{\beta_{n+1} + \frac{\alpha_{n+1}\gamma_{n+2}}{\beta_{n+2}  \ldots}}
\end{equation}
 viewed as an equation for $\omega_n$.  
For D3-brane, the differential equation can be put in the canonical form with two regular singular points and an irregular one  by  setting (with $L{=}\hbar{=}1$)
 \be
 r=1+2y+2\sqrt{y(1+y)}
 \ee 
 leading to
 \be
 Q(y) {\,=\,}{ 4y(y{+}1)\left[ 8 \omega^2(1{+}8 y(1{+}y)){-}(\ell{+}2)^2{+}1\right]
 {+}3  \over 16\, y^2(y{+}1 )^2} 
 \label{qhbar2}
 \ee
  which is the characteristic function of the ${\cal N}=2$ SYM with $SU(2)$ gauge group coupled to $(N_+,N_-)=(1,2)$ flavours  and parameters 
\bea
   u &=& (\ell+2)^2+ 2i  \omega -2  \omega^2 ~,~
   q= 8 i\, \omega\nn\\
   m_1 &=& m_2=0  ~,~  m_3=\ft12
\eea
There are two solutions around this point corresponding to $\sigma_+=\ft14$ and $\sigma_+=\ft34$. Comparing with the results coming from geodetic motion, and the large $q$ expansion of the Mathieu equation, one finds that the two choices reproduce QNM frequencies  with even and odd  overtones $n$'s respectively. 

The results coming from geodetic motion, SW techniques and numerical methods are plotted in figure \ref{fig-QNM-D3}. We find excellent agreement between the three methods even for small $\ell$, where the semi-classical geodetic approximation is not fully justified.
    
\section{Examples}
%
  {\bf Kerr-Newman BH}: The wave equation is separable into two ordinary differential equations describing radial and angular motion and depending on the frequency $\omega$, the separation constant $K$, the mass ${\cal M}$, angular momentum variable $a=J/{\cal M}$, the charge ${\cal Q}$ and an orbital mode $m_\phi$.  Both radial and angular equations match that of $SU(2)$ gauge theory with $(N_+,N_-)=(1,2)$ fundamentals with parameters 
\bea
u  &=&K^2 {+} \ft{1}{4}
      {-} \omega^2(5 r_+^2{+}r_-^2{+}2 r_{-} r_+) 
      {+} \omega ({\rm i} (r_+{-}r_-){-}4  a \zeta)\nn \\
q &=&- 2\,{\rm i}\, \omega\, ( r_+{-}r_-)
  \quad, \quad  m_1 = {{\rm i} \,\left[ 2\,a\,\zeta+\omega\,(r_-^2+r_+^2) \right]
          \over r_+{-}r_-   } \nn\\
m_2 &=&m_3={\rm i} (r_++r_-) \,\omega \quad, \quad y = {r-r_-\over r_+ - r_-} 
   \eea
for the radial part, with $\zeta=m_\phi-a\omega$,  and $r_\pm$ the inner and outer horizon radius $ r_\pm  =  {\cal M}  {\pm} \sqrt{  {\cal M}^2{-}a^2{-}{\cal Q}^2 }$.
For the angular part, using $y=(\cos\theta-1)/2$ one finds
\bea
   q^{\theta} &=& 4a \omega \quad, \quad u^{\theta}=K^2+\ft{1}{4} -2 a \omega(1{+}2 m_\phi) \nn\\
   m^{\theta}_1 &=& m_\phi \quad, \quad m^{\theta}_2=m^{\theta}_3=0
   \eea
   The extremal limit is obtained by sending $r_- \to r_+$. In this limit $q\to 0$ and $m_1\to \infty$ but the product $ \hat{q}=-q m_1$ stays finite. This leads to $(N_+,N_-)=(1,1)$ fundamentals with gauge coupling $ \hat{q}=  4\, \omega\, ( a\,\zeta+\omega \, r_+^2 )$.
  
 {\bf CCLP five-dimensional BHs}: CCLP metrics describe rotating solutions of Einstein-Maxwell theory in $d=5$, with mass ${\cal M}$, charge ${\cal Q}$, and angular momenta parameters $\ell_1$, $\ell_2$.  Introducing the variables $z=r^2$ and $\xi=\cos^2\theta$ the radial and angular wave equations can be mapped to $SU(2)$ gauge theory with $(N_+,N_-)=(0,2)$ flavours. The gauge/gravity dictionary for the radial wave equation reads
\begin{equation}
\begin{aligned}
u  &= \frac{1}{4}\left[1+K^2-\omega^2\left(4\mathcal{M}-{z}_-\right)\right]~,~
y = {z-z_-\over z_+ {-} z_-} \\
q &= \frac{\omega ^2 }{4}({z}_-{-}{z}_+)~,~
m_{1,2} =  -\frac{i}{2}\frac{\mathcal{L}_{\mathcal{M}}{\pm}\mathcal{L}_{\mathcal{Q}}}{\sqrt{z_+}{\mp}\sqrt{z_-}} 
\end{aligned}
\end{equation}
with 
\begin{align}
 z_\pm  &=  {\cal M}{-}\ft{\ell_1^2{+}\ell_2^2}{2} {\pm} \sqrt{ \left( {\cal M}{-}\ft{\ell_1^2{+}\ell_2^2}{2}\right)^2{ -}({\cal Q}{+}\ell_1 \ell_2)^2}\\
\mathcal{L}_{\mathcal{M}} &{=} \ell _2 m_{\psi }{+}\ell _1 m_{\phi }{-}2 \mathcal{M} \omega  ~,~
\mathcal{L}_{\mathcal{Q}} {= }\ell _1 m_{\psi }{+}\ell _2 m_{\phi }{+}\mathcal{Q} \,\omega \nn  
\end{align}
while the dictionary for the angular part ($y{\,=\,}-\xi$) reads
\begin{equation}
\begin{aligned}
q^\xi {=} \frac{ \omega ^2 }{4} (\ell _1^2{-}\ell _2^2)
~ , ~
m^\xi_{1,2} {=} \frac{m_\phi {\pm} m_\psi}{2} 
~ , ~
u^\xi {=} \frac{1{+}K^2{-}\omega^2 \ell _2^2}{4}
\end{aligned}
\end{equation}
 
{\bf D1D5 fuzzball:} Finally we consider a circular D1D5 profile with $a$ the radius and $Q_1=Q_5=L^2$. 
The wave equation now separates into equations of type (\ref{waveeq}) with $Q$'s functions matching that of $SU(2)$ gauge theory with $(N_+,N_-)=(0,2)$ fundamental hypermultiplets.
 The gauge/gravity dictionary for the radial variables is
\bea
q &{\,=\,}& {(\omega^2{-}P_y^2)\, a^2\over 4}   ~ , ~
  u {\,=\,}{K^2{+}1{-}(2L^2{-}a^2)(\omega^2{-}P_y^2) \over 4}   \nn\\
m_{1,2} &=& {L^2\over 2 a} (\omega\pm P_y) - {m_\phi \pm m_\psi  \over 2}  
~ , ~
y={\rho^2\over a^2} 
\eea
 while for the angular ones, using $y= {-}\cos^2 \theta$, one finds
\be
q = {a^2\over 4} (\omega^2{-}P_y^2) ~~ , ~~
u={K^2{+}1\over 4} ~~ , ~~
m_{1,2} = { m_\phi {\pm} m_\psi \over 2}
\ee

\noindent{{\bf{\em Acknowledgments.}}}

We would like to thank A.~Aldi, C.~Argento, G.~Bonelli, V.~Cardoso, G.~Di~Russo, D.~Fioravanti, M.~Firrotta, F.~Fucito, A.~Grassi, T.~Hikeda, M.~Mari\~no, P.~Pani, G.~Raposo, R.~Savelli, and Y.~Zenkevich 
for interesting discussions and valuable suggestions.

\bibliographystyle{utphys}
\bibliography{Ref}

\end{document}